\documentclass[cameraready]{Interspeech}

\usepackage{xcolor,colortbl}
\definecolor{loop}{HTML}{FF808E}
\definecolor{decision}{HTML}{8ab56c}
\definecolor{prior}{HTML}{c0d9e5}
\definecolor{estimate}{HTML}{fffea9}
\definecolor{ok}{HTML}{145908}
\usepackage[utf8]{inputenc}
\usepackage{multirow}
\usepackage{tabularray}
\usepackage{float}
\usepackage{graphicx}
\usepackage[export]{adjustbox}
\usepackage{algorithm}
\usepackage{algorithmic}
\usepackage{amsmath}
\usepackage{amssymb}
\usepackage{mathtools}
\usepackage{amsthm}
 
\newcommand{\Btargmax}{\mathop{\mathrm{B_t\text{-}argmax}}} 
 
\newcommand{\Bsqargmax}{\mathop{\mathrm{B^2\text{-}argmax}}} 
\usepackage{dirtytalk}
\usepackage[activate={true,nocompatibility},final]{microtype}
\SetTracking{encoding = *, shape = sc}{40}
\DisableLigatures{encoding = T1, family = tt*}

\usepackage{tikz}
\newcommand*\circled[1]{\tikz[baseline=(char.base)]{
            \node[shape=circle,draw,inner sep=0.6pt, font=\scriptsize] (char) {#1};}}

\theoremstyle{definition}

\usepackage{comment}
\title{A Generalized Formalism of Auto-Regressive Decoding for Speech Processing}

\author[orcid=0009-0009-7674-2303, correspondingauthor]{Julia}{Gachot}
\author[orcid=0000-0002-2355-0764]{Philipp}{Allgeuer}
\author[orcid=0009-0001-8734-2892]{Marie S.}{Bauer}
\author[orcid=0000-0003-1343-4775]{Stefan}{Wermter}

\address{Knowledge Technology, Department of Informatics, University of Hamburg, Germany}

\email{julia.gachot@uni-hamburg.de, philipp.allgeuer@uni-hamburg.de, marie.bauer@uni-hamburg.de, stefan.wermter@uni-hamburg.de}
\keywords{Auto-regressive (AR) models, Generation, Decoding, Deep Neural Networks (DNN)}

\begin{document}

\maketitle

\begin{abstract}
In speech processing, most state-of-the-art sequence prediction models rely on auto-regressive (AR) strategies to generate output sequences based on the raw predictions of the model. 
Despite their crucial role in the inference process, a comprehensive overview of AR strategies as a unified field is lacking, due largely to implicit and multiple definitions of next-token decoding.
This context complicates the choice, comparison, and evaluation of strategies, while creating inconsistencies in the characterization of approaches as auto-regressive or not.
We begin by setting explicit inclusion criteria for the field of AR search in speech processing, and derive a generalized theoretical framework to categorize and report on search strategies for neural models.
We show the capabilities of this formalism in simplifying the design of benchmarks centered around the decoding process, allowing for ablation studies that are focused on search strategies.
\end{abstract}

\section{Introduction}\label{sec:intro}
Across language modeling tasks, auto-regressive (AR) neural networks dominate the state-of-the-art for sequence prediction \cite{2602-prabhavalkar, 2602-tts-sota, 2602-sethiya, 2602-wiher}.
A model's generation strategy refers to the local search process towards composing a sequence, through an iterative scoring and updating of partial candidates.
These models are trained to estimate an auto-regressive score from an input and a prior, which represents the current decoding status.
At inference time, sequences are iteratively constructed to approximate a local maximum of an objective function, using the model's raw predictions.
The most commonly used objective function is Maximum a Posteriori (MAP). 
It is generally the one used for beam search \cite{2602-lowerre}, where the MAP estimates are combined to a breadth-first heuristic for the exploration of the search space. 
Other heuristics are present in literature, for example, that are tailored to improve performance on specific speech processing tasks.
In this direction, early strategies proposed variants on the objective function side, such as sampling \cite{2602-holtzman-temperature, 2602-holtzman, 2602-hewitt, 2606-delucia}, which introduces some randomness in the candidates.
More recently, from the modeling side, models called Non-Auto-Regressive (NAR) often maintain a notion of objective function, but propose a parallelized decoding process \cite{2602-xiaonar}.
The existence of these iterative sequence prediction methods, which deviates from the classic left-to-right next-token prediction, raises the question of where to draw the line between auto-regressive and other approaches.
New search strategies have also appeared in research papers around major developments in sequence modeling architectures, such as LLMs. 
The focus of these papers is often a task, a model, or a training method, while the inference process is treated more as an implementation detail.

As some of these decoding strategies are reused across tasks, they begin to exist in multiple versions (e.g., speculative sampling, NAR sequence generation), creating the need for a consistent way to place and compare methods with existing literature.
The lack of formalization of inference and training schemes as distinct aspects of modeling, given the drastically different questions and challenges they represent, has been identified in literature \cite{2512-bachmann}.
It is also reflected on the experimental side, where evaluating search strategies' behavior is difficult across speech processing tasks, and therefore rarely done \cite{2602-wiher, 2602-wang-nmt}.
To simplify making these benchmarks, we propose a generalized formalism to conceptualize and implement AR generation algorithms beyond specific tasks and models.
This framework aims to move the field away from the ubiquitous conceptualization of these approaches as `model likelihood maximizers' at inference time.
Instead, we define AR strategies for neural networks from the perspective of designing a recurrence relation that efficiently estimates sequence likelihood, updates candidates, and determines when and how to end the iterative process.
From this generalized and modular definition of the steps intrinsic to AR generation strategies, explicit criteria for a decoding method to be considered auto-regressive can be derived.
We examine cases that are traditionally viewed as being at the edge of the field, and show the capabilities of this framework in categorizing approaches in a systematic way, from the specificities of their structure.
Using the modularity of the proposed formalism, we also demonstrate how to isolate the contribution of each step to the overall performance by proposing an ablation study methodology for sequence prediction search strategies.

\section{Related Work}\label{sec:RW}
\subsection{AR decoding as a combinatorial optimization strategy}
Decoding strategies for neural sequence predictors constitute a subset of methods for discrete combinatorial optimization, in which heuristics are used to explore large structured solution spaces.
Several taxonomies propose an overview of these heuristics, but rarely cover machine learning approaches \cite{2602-stork}, or only for model optimization \cite{2602-rajwar}.
In sequence prediction, taxonomies tend to focus on AR architectures \cite{2602-tang-taxonomy}, and frameworks exist to evaluate the compositional capabilities of models \cite{2512-dziri}, but not their decoding strategies.
Without a systematic comparison framework in this rapidly evolving field, very similar metaheuristic algorithms are proposed under different names \cite{2602-rajwar}.
This also happens with certain subgroups of neural generation methods, such as the parallelization of the decoding process to decrease the number of inference steps, which can be found under many different names, although speculative decoding is the most commonly used \cite{2602-stern, 2602-ghazvininejad, 2512-sun, 2512-leviathan, 2602-santilli}.
One reason may be that presenting a `one-step edit' is often met with concerns from reviewers regarding the technical novelty of the approach, even for papers that later have a high impact \cite{2602-holtzman, 2602-vijayakumar}.
Within the framework proposed in this paper, we compose a general structure of AR decoding as a modular set of steps to simplify locating and characterizing contributions within the structure in both existing and future algorithms.

Even beam search usually does not refer to the original algorithm anymore \cite{2602-lowerre}, but has been adapted with variations for neural networks, especially regarding their termination condition.
Taking for granted the contents of searches creates a risk of insufficient reporting, which is documented even beyond the case of generation with neural models \cite{2602-barley}.
By moving away from presenting strategies as an indivisible block, our framework clarifies the core elements to document when presenting or using a strategy, detailed in Sec.~\ref{sec:report}.
Regarding these steps, some formalisms in the literature describe beam search as several steps being repeated \cite{2602-beraldi}, but they do not detail how the estimation could be performed with a sequence prediction model.
For neural models, the formalism of Best-First Beam Search is general enough to recover some beam search variants, but it is not designed to encompass stochastic methods \cite{2506-meister}.
Our formalism encompasses a wider diversity of approaches and is oriented specifically toward neural models.

\subsection{Current categorizations of AR strategies}
Without an existing taxonomy, the scope of surveys on AR strategies can be used as a proxy for how strategies are currently grouped in the literature, as well as for the open questions they reflect.
AR decoding strategies are often surveyed in a task-specific manner, highlighting a strong task-method association for certain tasks.
In automatic speech recognition (ASR), end-to-end speech-to-text translation, and grammatical error correction (GEC), beam search and its variants are the expected strategies \cite{2602-prabhavalkar, 2602-sethiya, 2602-bryant}.
In machine translation (MT), beam search also appears as the reference \cite{2506-leblond, 2602-freitag}, although some works suggest that stochastic methods could also be promising \cite{2602-eikema}.
Text-To-Speech (TTS) is a task where NAR models have been omnipresent over the last five years, with less attention given to search strategies than in some NLP tasks \cite{2602-kaur, 2602-tts-sota}.
In the language model-based subset of TTS, sampling with temperature is classically used, like VoiceCraft \cite{2602-voicecraft}.
For LLM-based generation task benchmarks, however, beam search is mostly used as a baseline and is often systematically evaluated against stochastic methods \cite{2602-shi-survey, 2602-wiher, 2602-welleck}.
This has led to the widespread view of deterministic and stochastic strategies as two distinct classes of methods, due to the benefits of randomness in introducing diversity in generated outputs \cite{2602-meister-locally, 2602-arora, 2602-vijayakumar}.
We argue that this intuition is no longer representative of the field.
Some non-stochastic methods generate diverse outputs by penalizing repetitions \cite{2503-tu-tts}, or by performing a sampling step that, while not involving randomness in the process, would exclude them from MAP strategies if interpreted strictly \cite{2602-beraldi}.
As a consequence, findings about AR strategies across tasks are rarely explored, and the papers that inspect it demonstrate vast performance differences even within text generation \cite{2602-wiher, 2602-josifoski, 2602-ippolito}.
In this paper, by adopting a modular view of these approaches, we propose a formalism that categorizes these searches more reliably, keeping the same structure regardless of whether the search is deterministic or stochastic, or tested for a specific task.

\section{Generalized Auto-Regressive Formalism} \label{sec:form}
This framework is intended for tasks that can be formalized as a discrete compositional optimization problem, and more precisely a stochastic integer problem under probabilistic constraints (SIPC) \cite{2602-beraldi}. 
In SIPC, both random variables and decision variables are integers, which is why outputs are often represented as an ordered one-dimensional sequence of integers or tokens.

\subsection{Assumptions and inclusion criteria}\label{sec:hyp}
We define a generation approach as a tuple $(\mathcal{M}, \ g_{AR})$, where $\mathcal{M}$ is a neural model, and $g_{AR}$ its inference algorithm. 
In the following, we call this tuple an auto-regressive predictor, provided it fits the following criteria.

\noindent\textbf{Model assumptions.} The model is trained to estimate conditional probabilities over $\mathcal{A}$, a finite alphabet or set of tokens.
During inference, the notion of decoding status is leveraged to estimate the conditional probability through the updating of an internal state or a prior. 
The model's predictions belong to $\mathcal{A}^*$, the Kleene closure of its alphabet.

\noindent\textbf{Decoding assumptions.} The generation process $g_{AR}$ is an iterative local maxima search, designed to be tight \cite{2602-du-lee}. 
An objective function (e.g., MAP) is used at least once throughout the process to inform the updating of a finite set of candidates sampled from $\mathcal{A}^*$.

\subsection{General formalism}\label{sec:choices}
Based on the assumptions in Section~\ref{sec:hyp}, we now define the core steps of an auto-regressive predictor's inference algorithm $g_{AR}$.
In the general SIPC formalism, at each iteration $t$ of the decoding process, a set of $B_t \in \mathbb{N}_+$ candidate sequences $\mathbf{Y}_{t} = \{\mathbf{y}^{b}\in \mathcal{A}^*| \ b \in 1...B_t\}$ is updated.
The specificity of using SIPC strategies with neural models is that estimating the conditional probability requires a notion of a prior, denoted as a set $\mathbf{Z}_t$, to keep track of the decoding process using various sources of information.
In the simplest cases, $\mathbf{Z}_t$ coincides with the set $\mathbf{Y}_{t}$ of candidates resulting from the $t-1$ inference step, but an alternative state variable is used by some architectures.

While the nature of the prior $\mathbf{Z}_t$ is often evident for a specific algorithm, proposing a general definition for it is one of the key levers of our formalism for enabling a systematic comparison between models and tasks.
Additionally, examining if and how a prior is defined and updated throughout the generation process is a good proxy to distinguish auto-regressive (AR) strategies from other local search methods.
In the same spirit, instead of defining an iteration around the evaluation of an objective function, we outline a modular structure for AR strategies that generalizes across speech processing tasks.
We formalize an iteration $t$ as a function $f_{(\mathcal{M}, \ g_{AR})}^t(\mathbf{Y}_{t}, \mathbf{Z}_{t})$ that returns an updated prior $\mathbf{Z}_{t+1}$ and set of candidate sequences $\mathbf{Y}_{t+1}$ from the past step output and prior.
\begin{align}
    \label{alg:iter}
    (\mathbf{Y}_{t+1}, \mathbf{Z}_{t+1}) = f_{(\mathcal{M}, \ g_{AR})}^t(\mathbf{Y}_{t}, \mathbf{Z}_{t}), \ \text{with} \ \mathbf{Y}_{t+1} \ne \mathbf{Y}_{t}
\end{align}
Based on the assumptions of Sec.~\ref{sec:hyp}, once all of the iterations $f_{(\mathcal{M}, \ g_{AR})}^t$ are complete, each of the following four steps has happened at least once:

\noindent\textbf{\circled{1} Estimation.} For each candidate, the model $\mathcal{M}$ is assumed to estimate a probability mass function (PMF).
The output of this step is $P_t$, a PMF conditioned on $\mathbf{Y}_{t}$ and $\mathbf{Z}_{t}$ .
\begin{align}
\label{eq:pmf}
\begin{split}
    P_{t+1}: \mathcal{A}^* &\longrightarrow [0,1] \\
    \mathbf{a} & \longmapsto p(\mathbf{a} \ | \ \mathbf{Y}_{t}, \mathbf{Z}_{t})
\end{split}
\end{align}
This is often implemented by performing the forward pass of an input $\mathbf{x}$, post-processing the resulting logits as necessary, and applying a softmax.
In this classic case, the PMF is given by
\begin{align}
    \label{eq:fwd}
    P_{t+1} = \text{softmax}(\mathcal{M}(\mathbf{x}, \mathbf{Z}_t)).
\end{align}

\textbf{\circled{2} Decision.} When a generation approach is auto-regressive, the model estimation should intervene in refining the candidate sequence sets.
By evaluating an objective function $f_{obj}$, the model's PMF estimates from eq.~\ref{eq:pmf} are aggregated and interpreted.
The highest scoring output according to $f_{obj}$ can then be used to construct $\mathbf{Y}_{t+1}$.
If a sequence is generated from left-to-right, the decision step consists of evaluating the $B_t$ highest scoring sequence continuations $\mathbf{a}_{t+1}\in\mathcal{A}^*$ to the elements of $\mathbf{Y}_t$.
Based on this objective $f_{obj}$, the candidate sequences become
\begin{align}
    \label{eq:bargmax}
    \mathbf{Y}_{t+1} = \Btargmax_{\mathbf{y}\in \mathcal{A}^*} f_{obj}(\mathbf{y}).
\end{align}
The following is expressed with that idea in mind for clarity, but can easily be adapted to other cases, such as parallelized generation strategies that do not adopt that implementation.
With a Maximum A Posteriori (MAP) objective, the general decision step from eq.~\ref{eq:bargmax} becomes
\begin{align}
    \label{eq:map}
    \mathbf{Y}_{t+1} = \Btargmax_{\mathbf{y}\in \mathcal{A}^*} \sum_{\mathbf{a}_t\in \mathbf{y}} \log  \left(P_{t}(\mathbf{a}_t)\right).
\end{align}

\textbf{\circled{3} Update of prior.} 
This last step is based on the assumption that a generation approach is considered auto-regressive if it includes a mechanism to keep track of the past inference steps.
In the simplest (and most common) case, $\mathbf{Z}_t$ is a set of sequences from $\mathcal{A}^*$.
At each decoding iteration $t$, we have that the prior passed for the estimation step is exactly the set of candidates $\mathbf{Y}_{t}$, meaning $\mathbf{Z}_{t} \gets \mathbf{Y}_{t}$.
Otherwise, the relation can have a more complex expression, for example, if the prior is based on the model's internal state.

\textbf{\circled{4} Termination test.}
Finally, a boolean termination condition must ensure that, after a finite number of iterations, the decoding process terminates. 
At a given iteration $t$, that is tested using
\begin{align}
    f_{term}: (\mathcal{A}^{*})^{B_t}  &\to \{0,1\} \\
            \mathbf{Y}_{t} & \mapsto \begin{cases*}
                    1 & \text{if termination condition} \\
                    0 & \text{otherwise,} 
                 \end{cases*}
\end{align}
where $f_{term}$ denotes the function that tests the candidate $\mathbf{Y}_{t}$ for decoding termination.
\begin{figure}[t]
  \centering
  \includegraphics[width=\linewidth]{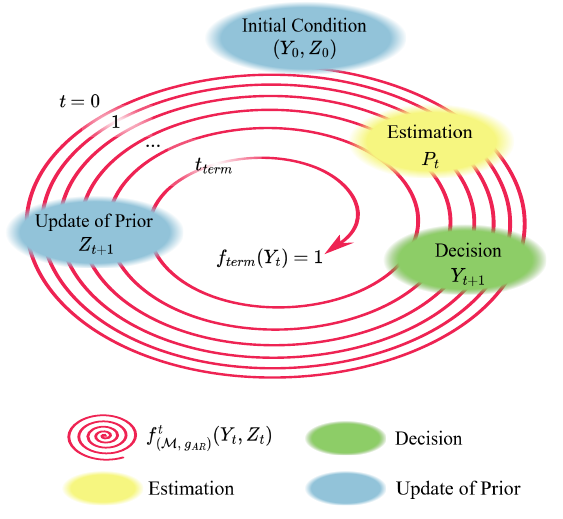}
  \caption{Structure of the most common local search process for sequence generation with neural models.
  The formulation of the iteration $f_{(\mathcal{M}, \ g_{AR})}^t(\mathbf{Y}_{t}, \mathbf{Z}_{t})$ does not depend on the iteration step $t$, which is true for beam search and most of its variants. 
  At each iteration, an estimation, a decision, and an update of the prior are performed until reaching a satisfying approximation at the center.
  The termination condition is defined to reflect what an acceptable output is for the algorithm at hand.}
  \label{fig:classic}
\end{figure}

\subsection{Reporting design choices}\label{sec:report}
Within the framework defined in this paper, defining an auto-regressive decoding strategy $g_{AR}$ consists of making a set of design choices governing a recurrence relation.
This begins with specifying an \textbf{initial condition}, i.e., the contents of $\mathbf{Y}_{0}$ and $\mathbf{Z}_{0}$. 
Then the \textbf{recurrence relation} of eq.~\ref{alg:iter} can either be the same at each iteration, meaning defining $f_{(\mathcal{M}, \ g_{AR})}^0$ is enough, or if the recurrence relation evolves throughout the decoding, each iteration expression is required.
Reporting on an iteration $t$, means defining the \textbf{estimation}, \textbf{decision}, and/or \textbf{update of prior} steps within $f_{(\mathcal{M}, \ g_{AR})}^t$, as well as if it includes a \textbf{termination test}.

\subsection{Formalizing classic search strategies}\label{sec:recovering}
We demonstrate the capability of our framework in encompassing both beam search and sampling by following the reporting guidelines of Sec.~\ref{sec:report}.

\subsubsection{Beam search}\label{sec:beam}
A specificity of beam search is its beam size, the fixed number $B \in \mathbb{N}_+$ of candidate sequences in any $\mathbf{Y}_t$.
The \textbf{initial condition} in beam search depends on the training strategy of $\mathcal{M}$, the chosen model, e.g., BOS or prompt tokens.
Regarding the \textbf{recurrence relation}, in beam search, all iterations follow the same template, illustrated in Fig.~\ref{fig:classic}, and contain an estimation and decision steps followed by a termination test. If the termination test is negative, the iteration ends with an update of the prior, or nothing if the test is positive. 
Each \textbf{estimation} consists of applying the model $\mathcal{M}$ forward pass to the input sample ($\mathbf{x}$) and prior at $t$ ($\mathbf{Y}_t$) to generate logits. The softmax function is applied to the logits and yields the PMF for step $t+1$, as described in \circled{1}. The \textbf{decision} is based on the Maximum A Posteriori objective function of eq.~\ref{eq:map}.
Each \textbf{update of the prior} consists of setting $\mathbf{Z}_{t+1} \gets \mathbf{Y}_{t+1}$. 
The \textbf{termination} of beam search is also dependent on the model; for example, the first appearance of an End-Of-Sequence (EOS) token within the set of candidates can be a termination test. 
Once the test is positive, at $t_end$, the most likely sequence in $\mathbf{Y}_{t_{end}}$, is returned as a final output.

\subsubsection{Temperature sampling} 
In sampling with temperature \cite{2602-holtzman-temperature, 2602-ackley}, $B$ candidates are updated at each iteration.
Similarly to Sec.~\ref{sec:beam}, the order and content of steps within an iteration remain identical throughout the decoding.
For a given model, switching from beam search to sampling has no influence on the definition of the initial condition, termination and update of prior.
Within our framework, the differences between these two methods are restricted to the estimation and decision steps.
For the \textbf{estimation}, temperature rescaling introduces a slight difference in the estimation step's eq.~\ref{eq:fwd}.
It becomes $P_{t+1} = \text{softmax}(f_T(\mathcal{M}(\mathbf{x}, \mathbf{Z}_t)))$ for $f_T$ the temperature sampling function, applied to the raw logits.
At the \textbf{decision} step, the aggregated probability estimates are filtered to keep $B^2$ sequences ranked using MAP, like in eq.~\ref{eq:map}, except $\Btargmax$ becomes $\Bsqargmax$. Finally, $B$ sequences from the resulting set are randomly sampled to compose $\mathbf{Y}_{t+1}$.

\section{Discussion}\label{sec:discuss}
\subsection{Edge cases} \label{sec:edge}
Presented with sequence prediction methods that deviate from the classic left-to-right next-token prediction paradigm, we demonstrate how our proposed formalism draws the line between auto-regressive and other approaches.
We look into ten methods over various speech processing tasks, published between 2018 and 2025.
Among them, half propose a speculative decoding strategy \cite{2602-stern, 2602-ghazvininejad, 2512-sun, 2512-leviathan, 2602-santilli}, while the others present non-AR methods \cite{2602-gu-nar, 2603-chen-narasr, 2603-peng-ttsnar, 2602-xu-narst, 2603-yang-nartts}.
The principle for inclusion is to consider a generation approach to be auto-regressive if both the model and decoding assumptions defined in Sec.~\ref{sec:hyp} hold.
Among the ten papers, based on the \textbf{decoding assumptions}, only one approach is excluded for not formulating the task as an SIPC \cite{2603-peng-ttsnar}.
The \textbf{model assumptions} hold for all speculative decoding methods as well as, perhaps surprisingly, two approaches from the non-AR group \cite{2603-chen-narasr, 2603-yang-nartts}.

Based on these assumptions, models that do not estimate a conditional probability are excluded, even if a mechanism is used to constrain the coherence of sequence \cite{2602-gu-nar}.
However, a diversity of approaches is included, even when using novel variables as prior for their conditional probability estimation step.
These priors often differ from the usage of those employed in traditional next-token predictors, or their construction from past candidate sequence sets $\mathbf{Y}_t$, but the overall structure of their generation process remains analogous.
As a result, our framework can be used to compare traditional methods with algorithms that do not adopt the default left-to-right decoding process \cite{2602-ghazvininejad, 2603-chen-narasr}, or do not constrain how far ahead a prior can influence the conditional probability \cite{2602-stern, 2603-yang-nartts}.
In addition to relaxing the aforementioned \textit{left-to-right} and \textit{next-token} aspects, as long as the generation is conditional on a decoding status variable, even using past candidate sequence sets is not mandatory to update a prior  \cite{2512-sun, 2602-xu-narst}.
In addition to methods that can intuitively be identified as auto-regressive, our proposed framework includes search approaches that follow a similar structure, making them comparable to the rest of the field.

\subsection{Categorization}\label{sec:categ}
As discussed in Sec.~\ref{sec:choices}, the proposed framework is designed to describe and interpret generation strategies as local search by clarifying their recurrence relation mechanism, as described in eq.~\ref{alg:iter}.
Each of the design choices \circled{1} to \circled{4} that constitute this relation can also be used to compare methods from the standpoint of their design choices and structure.
Most search strategies adopt an iteration formulation that does not change over time as described in Fig.~\ref{fig:classic}, meaning $f_{(\mathcal{M}, \ g_{AR})}^0$ is reused at each iteration $t$.
In that case, the order and individual definition of steps are enough to describe the generation process entirely, from which an immediate and systematic step-by-step comparison can be made to compare generation strategies.
Moreover, most search strategies for neural networks share steps with beam search or other traditional methods, meaning the nature of the proposed changes can be located at specific design choices.
Note that various steps propose novel estimation steps  \cite{2512-leviathan, 2602-xu-narst, 2602-holtzman-temperature}, decision steps \cite{2602-stern, 2603-yang-nartts, 2602-vijayakumar, 2503-tu-tts}, or update the decoding prior novel ways \cite{2602-ghazvininejad, 2512-sun, 2602-santilli}.
These categories propose a new angle on how to group tasks, to complement the classic paradigms of restricting to a task, or to deterministic or stochastic strategies.

When constructing a benchmark of search strategies that are compatible with a given model, our framework groups methods that share a similar location or principle to optimize for speed, diversity, and so on.
For instance, when maximizing diversity, a TTS strategy based on the MAP objective \cite{2503-tu-tts} and a stochastic decoding method for text generation \cite{2602-vijayakumar} may not seem relevant to compare at first glance. 
However, within our framework, these strategies share most of their structure, and their contributions can both be located and understood as decision step variants.
This categorization by steps holds beyond the traditional tasks and objective functions' scopes, which allows to compare and to draw new parallels between existing strategies.

\subsection{Prospective ablations for AR decoding}\label{sec:mix}
When proposing a new model architecture, it is a common practice to perform ablation studies.
For non-AR architectures, it often consists of evaluating the individual contribution of a score or architecture change to inference speed or output quality \cite{2602-gu-nar, 2603-peng-ttsnar, 2603-yang-nartts}.
To evaluate generation strategies, similar experiments are sometimes performed by varying hyperparameter values to limit the effect of a step, such as the termination condition \cite{2512-sun}.
Within the formalism from Sec.~\ref{sec:choices}, the added modularity in the structure can be used for assessing the respective contributions of novel estimation, decision, update of prior, or termination condition to the overall performance.
Since these steps are structural components, instead of completely canceling a step entirely, the ablation is approximated by switching one or several steps with a baseline equivalent (e.g., the same step from beam search).
The more the results are impacted by the replacement of a step, the more it contributes to overall performance, similarly to classic ablation studies.

This can help create links with non-AR models ablation studies, where new scores or architecture changes are often created as a way for the decoding process to be parallelized within the model.
This modular view of searches can also open new research directions, having identified which elements of a search make the most effect.
It creates the possibility to take advantage of several contributions of existing searches in certain cases, optimizing for more than one goal, including speed, diversity, or accuracy.
For example, a search strategy that proposes a score at step \circled{3} to increase diversity could be combined with an estimation method \circled{1} that speeds up the decoding.
This addresses the need for more efficient and powerful searches, while keeping track of the provenance of each block, rather than by presenting each search as an entirely new and indivisible algorithm.

\section{Conclusion}
In this paper, we propose a framework that sets the recurrence relation at the core of reporting on generation strategies, including systematic inclusion criteria for a search to be auto-regressive or not (Sec.~\ref{sec:edge}).
It offers a new way to compare searches from the location and mechanism of their contributions within the search process (Sec.~\ref{sec:categ}).
Moreover, by shifting the focus from the comparison with searches that make similar claims or are used for the same task, our framework offers a new angle to include comparable auto-regressive predictors and compose high quality benchmarks.
Finally, we transpose the notion of ablation studies to decoding strategies, and highlight possible research directions when designing new generation algorithms by optimizing for several aspects at the same time (Sec.~\ref{sec:mix}).


\section{Acknowledgments}
The authors gratefully acknowledge funding from Horizon Europe, under the MSCA grant agreements 101072488 (TRAIL), 101168792 (SWEET) and 101226624 (GREET).
This work was also supported by the German Research Foundation (DFG), for the project 551629603 (LUMO).

\section{Generative AI Use Disclosure}
No generative model was used in the writing of this article.

\bibliographystyle{IEEEtran}
\bibliography{mybib}
\end{document}